\documentstyle[pra,aps]{revtex}
\draft

\begin{document}

\title{Four-state $N$-atom as a simple theoretical model for
electromagnetically induced absorption}
\author{A. V. Taichenachev, A. M. Tumaikin and V. I. Yudin}
\address{Novosibirsk State University,
Novosibirsk 630090, Russia
}
\date{\today}
\maketitle

\begin{abstract}
A simple theoretical model describing the positive sign of
subnatural-width absorption resonances in the recent experiment of
Akulshin and co-workers (Phys. Rev. A, {\bf 57}, 2996 (1998)) is
proposed. An analytical expression for the linear response
to the weak probe field is found in the low-saturation limit with
respect to the control field. It is shown that the positive sign of
subnatural resonance is caused by the spontaneous transfer of the
light-induced coherence from the excited level to the ground one.
\end{abstract}


\section{Introduction}
As is well-known the nonlinear interference effects in the resonant
atom-light interaction can lead to the electromagnetically induced
transparency (EIT) of atomic medium \cite{EIT} as well as to other
interesting phenomena \cite{CPT}.
The key point of all these phenomena is the light-induced coherence
between atomic levels, which are not coupled by dipole transitions.
Recently Akulshin and co-workers have observed
subnatural-width resonances in the absorption on the $D_2$ line
of rubidium vapor under excitation by two copropagating optical waves
with variable frequency offset \cite{EIA}. Surprisingly, that apart
from EIT-resonances with negative sign, they have detected positive
resonances termed in ref.\cite{EIA} as electromagnetically induced
absorption (EIA). Basing on the experimental results and numerical
calculations, authors of ref.\cite{EIA} have deduced \cite{EIA2} that
EIA occurs in a degenerate two-level system when the three conditions
are satisfied:  i) The excited-state total angular momentum
$F_e=F_g+1$.  ii) Transition $F_g \rightarrow F_e$ is closed.  iii)
The ground state is degenerate $F_g > 0$.

In the present paper, motivated by the absence of EIA-resonances in
three-state $\Lambda$ and $V$-systems, we propose a simple
theoretical model for EIA -- four-state $N$-atom.
Namely, we consider an atom with four states $| i\rangle,\;\;
i=1\ldots 4$. The odd states $| 1\rangle$ and  $| 3\rangle$ are
degenerate and belong to the ground level, while the even states
$| 2\rangle$ and  $| 4\rangle$
(also degenerate) form the excited level. All optical transitions
$| odd\rangle \rightarrow | even\rangle$ are permitted except for
$| 1\rangle \rightarrow | 4\rangle$, that is forbidden. The
control field with frequency $\omega_1$ drives the transitions
$| 1\rangle \rightarrow | 2\rangle$ and $| 3\rangle \rightarrow |
4\rangle$. The weak probe at $\omega_2$ is applied to the $| 3\rangle
\rightarrow | 2\rangle$ transition.  An analytical expression for
the probe absorption as a function of frequency offset
$\omega_1-\omega_2$ is found. It is shown that EIA appears due to the
spontaneous transfer of the light-induced low-frequency coherence
from the excited level to the ground one. Obviously, both $\Lambda$
and $V$-systems do not describe such a process.  The sign of the
subnatural resonance depends on the branching ratio constant $0 \leq
b\leq 1$ and becomes positive for closed transition $b=1$. Velocity
averaging in the case of Doppler broadening is briefly discussed.

It is worth noting, effects of the
spontaneous coherence transfer on nonlinear resonances in the probe
field spectroscopy were first considered from the general point of
view by S.  G.  Rautian \cite{cohcasc}.

\section{Formulation of the problem}
Let us consider the resonant interaction of a bichromatic light
field:
\begin{equation} \label{field}
{\bf E}({\bf r},t) =
{\bf E}_{1} \exp[-i\omega_1 t+i({\bf k_1}{\bf r})] +
{\bf E}_{2} \exp[-i\omega_2 t+i({\bf k_2}{\bf r})] + c.c.
\end{equation}
with a four-state atom. This atomic system has four states
$| i\rangle,\;\; i=1\ldots 4$  (see in fig.1). The two odd states
$| 1\rangle$ and  $| 3\rangle$ are
degenerate and belong to the ground level with zero energy and zero
relaxation rate. The even states are also degenerate and form the
excited level with the energy $\hbar \omega_0$ and relaxation rate
$\Gamma$. We assume that among optical transitions between the ground
and excited levels $| odd\rangle \rightarrow | even\rangle$ the
transition $| 1\rangle \rightarrow | 4\rangle$ is forbidden due to
some selection rule (for instance, with respect to the momentum
projection).  Let the first term in eq.(\ref{field})
(will be refered as a control field) is sufficiently larger
than the second one, which is a probe field.
The control field with the vector amplitude ${\bf E}_1$ and
frequency $\omega_1$ drives simultaneously two transitions
$| 1\rangle \rightarrow | 2\rangle$ and $| 3\rangle \rightarrow |
4\rangle$. The weak probe filed ${\bf E}_2$ at the
frequency $\omega_2$ induces the $| 3\rangle\rightarrow | 2\rangle$
transition. In the rotating frame the Hamiltonian for the free
atom reads
\begin{equation} \label{ham0}
\widehat{H}_0=\hbar \delta_1 |1\rangle\langle 1| +
\hbar \delta_2 |3\rangle\langle 3|+
\hbar (\delta_2 -\delta_1)|4\rangle\langle 4| \,,
\end{equation}
where $\delta_q = \omega_q - \omega_0 -({\bf k}_q {\bf v})$
($q=1,2$) are the detunings including the Doppler shifts.
Using the rotating wave approximation, we write the atom-field
interaction Hamiltonian in the form
\begin{equation} \label{hamAF}
\widehat{H}_{AF}=\hbar \Omega_1 \widehat{Q}_1 +
\hbar \Omega_2 \widehat{Q}_2+h.c. \,.
\end{equation}
Here $\Omega_q$ are the corresponding Rabi frequencies and
the operators $\widehat{Q}_q$ are given by
\begin{eqnarray} \label{Qop}
\widehat{Q}_1 &=& A |2\rangle\langle 1| +
|4\rangle\langle 3|\nonumber\\
\widehat{Q}_2 &=& B |2\rangle\langle 3|\,,\;\;\; A^2+B^2=1 \,,
\end{eqnarray}
where the real numbers $A$ and $B$ govern the relative amplitudes of
transitions in the model under consideration.

In the case of the pure radiative relaxation the optical Bloch
equations for the atomic density matrix $\widehat{\rho}$ read
\begin{equation} \label{OBE}
\frac{d}{dt}\widehat{\rho}+
\frac{i}{\hbar}\left[\widehat{H}_{0}+\widehat{H}_{AF},\,\widehat{\rho}\right]
+\frac{1}{2}\Gamma \left\{\sum_{q=1,2}\widehat{Q}_q \widehat{Q}^{\dagger}_q,
\,\widehat{\rho}\right\}
- b \Gamma \sum_{q=1,2}\widehat{Q}^{\dagger}_q \widehat{\rho}\widehat{Q}_q
=\widehat{R} \, ,
\end{equation}
where the third term on the l.h.s. has a structure of
anticommutator and describes the radiative damping of the
excited-level populations and optical coherences. The last term on
the l.h.s.  corresponds to the transfer of the populations and
low-frequency coherences from the excited level to the ground one
under the spontaneous emission. The branching ratio constant $b$ can
vary between $0$ and $1$ and governs a degree in which the atomic
system is open. For example, $b=1$ means that the considered
transition is closed.  The r.h.s. of eq.(\ref{OBE}) is a source
describing possible external pumping of levels.

\section{Linear response to the probe field}
As is well-known, under the stationary conditions the intensity
${\cal I}_2$ of the probe field propagating through atomic
medium obeys the equation
$$
\frac{d {\cal I}_2}{d z}=-\hbar \omega_2 {\cal N}\,
( iB\Omega^{*}_2 \langle\rho_{23}\rangle_v+c.c ) \; ,
$$
where the $z$-axis is directed along ${\bf k}_2$, ${\cal N}$ is
the atomic density, and $\langle\ldots\rangle_v$ means averaging
over the atomic velocity. Thus, the probe field absorption is
proportional to the real part of the product $i\Omega^{*}_2
\rho_{23}$.  The steady-state off-diagonal element $\rho_{23}$ can be
written as
\begin{equation} \label{opcoh}
\rho_{23}=[\Gamma/2-i\delta_2]^{-1}
\{-iB\Omega_2(\rho_{33}-\rho_{22})
-iA\Omega_1\rho_{13}+i\Omega_1\rho_{24}\} \,.
\end{equation}
The last two terms in the curly brackets in eq.(\ref{opcoh})
(proportional to $\Omega_1$)
describe modifications of the absorption due to the light-induced
low-frequency coherences. It is well-known that in both $\Lambda$ and
$V$ systems the coherences $\rho_{13}$ and $\rho_{24}$ have such
phases that the absorption of the probe field is reduced at the
two-photon resonance $\delta_2=\delta_1$. We note that these two
terms in eq.(\ref{opcoh}) have opposite signs. In the case under
consideration an additional term in equation for $\rho_{13}$ arises
from the spontaneous transfer $\rho_{24} \rightarrow \rho_{13}$
($d\rho_{13}/dt =\ldots + b A \Gamma \rho_{24}$). Hence, the phase of
$\rho_{24}$ giving the transparency through the term
$i \Omega_1 \rho_{24}$ can give the icrease of absorption through
the term $-i A \Omega_1 \rho_{13}$.
For the sake of clarity, in the following analyses we use two
approximations: i) The first  order in the probe amplitude
$\Omega_2$; and ii) The low-saturation limit for the control field,
i.e. $\Omega_1 < \Gamma$. In this case instead of eq.(\ref{opcoh}) we
write \begin{equation} \label{linresp}
\rho^{(1)}_{23}=[\Gamma/2-i\delta_2]^{-1}
\{-iB\Omega_2 \rho^{(0)}_{33}
-iA\Omega_1\rho^{(1)}_{13}\} \,,
\end{equation}
where the index over $\rho^{(n)}$ means that this element is taken in
the $n$-th order on $\Omega_2$.  The equation (\ref{linresp}) should
be completed by the following equations for the first-order
coherences:
\begin{eqnarray} \label{1order}
i(\delta_1-\delta_2)\rho^{(1)}_{13} &=&  iB\Omega_2\rho^{(0)}_{12}
-iA\Omega_1^{*} \rho^{(1)}_{23}
+i\Omega_1\rho^{(1)}_{14}
 + bA\Gamma\rho^{(1)}_{24} \nonumber\\
\left[\Gamma+i(\delta_1-\delta_2)\right] \rho^{(1)}_{24} &=&
-iB\Omega_2\rho^{(0)}_{34} - iA\Omega_1 \rho^{(1)}_{14}
+ i\Omega_1^{*}\rho^{(1)}_{23} \nonumber \\
\left[\Gamma/2+i(2\delta_1-\delta_2)\right]\rho^{(1)}_{14} &=&
i \Omega_1^{*} \rho^{(1)}_{13} \\
\rho^{(0)}_{12} =
\frac{iA\Omega_1^{*}\rho^{(0)}_{11}}{\Gamma/2+i\delta_1}\,; &&
\rho^{(0)}_{34} =
\frac{i\Omega_1^{*}\rho^{(0)}_{33}}{\Gamma/2+i\delta_1} \;.\nonumber
\end{eqnarray}
Here we assume that the term $\widehat{R}$ in
eq.(\ref{OBE}) is diagonal and, consequently, gives
contributions into eqs.(\ref{1order}) implicitly through the
zero-order populations $\rho^{(0)}_{ii}$ only.
From eqs.(\ref{1order})  one can get
the coupled equations for the low-frequency coherences:
\begin{eqnarray} \label{lfcoh}
\left[\frac{|A\Omega_1|^2}{\Gamma/2-i\delta_2}+
\frac{|\Omega_1|^2}{\Gamma/2+i(2\delta_1-\delta_2)}+
i (\delta_1-\delta_2)\right] \rho^{(1)}_{13}
- b A \Gamma \rho^{(1)}_{24}& = &
-\frac{AB \Omega_2 \Omega_1^{*}}{\Gamma/2-i\delta_2}\rho^{(0)}_{33}
-\frac{AB \Omega_2 \Omega_1^{*}}{\Gamma/2+i\delta_1}\rho^{(0)}_{11}
 \nonumber \\
\left[\Gamma+i(\delta_1-\delta_2)\right] \rho^{(1)}_{24}
-\left\{\frac{A|\Omega_1|^2}{\Gamma/2-i\delta_2} +
\frac{A|\Omega_1|^2}{\Gamma/2+i(2\delta_1-\delta_2)}\right\}
\rho^{(1)}_{13} &=&
\left\{\frac{B \Omega_2
\Omega_1^{*}}{\Gamma/2-i\delta_2} +
\frac{B \Omega_2 \Omega_1^{*}}{\Gamma/2+i\delta_1}\right\}
\rho^{(0)}_{33} \,.
\end{eqnarray}
The right-hand sides of eqs.(\ref{lfcoh}) are the field interference
terms describing the creation of $\rho^{(1)}_{13}$ and
$\rho^{(1)}_{24}$. In the square bracket of the first line the field
broadening and optical shifts of the ground-level states are present.
Equations (\ref{lfcoh}) are not independent due to the second terms
on l.h.s. of both lines, which correspond to the spontaneous and
induced coherence transfer between levels.  In the low-saturation
limit the excited-level coherence $\rho^{(1)}_{24}$ enters into the
equation for the ground-level coherence $\rho^{(1)}_{13}$ through the
term describing the spontaneous coherence transfer. As it can be
seen, this process leads to changes in the position, width and
amplitude of nonlinear resonances connected with the low-frequency
coherence.  Since in the present paper we are interesting in the
subnatural-width resonance,
the anzatz $|\delta_1-\delta_2| \ll \Gamma$ is relevant.
Using eqs.(\ref{lfcoh}), we can eliminate the low-frequency
coherence in eq.(\ref{linresp}) and arrive at the final result for
the linear response:
\begin{eqnarray} \label{final}
\rho^{(1)}_{23}&=& \frac{-iB\Omega_2}{\Gamma/2-i\delta_2}
\left\{
\rho^{(0)}_{33} +
\frac{(b-1)\rho^{(0)}_{33}|A\Omega_1|^2}
{|A\Omega_1|^2(1-b)+
|\Omega_1|^2(1-bA^2)\frac{\Gamma/2-i\delta_2}{\Gamma/2+i\delta_1}
+ i(\delta_1-\delta_2)(\Gamma/2-i\delta_2)}+ \right.
\nonumber\\
&+& \left.
\frac{(b\rho^{(0)}_{33}-\rho^{(0)}_{11})|A\Omega_1|^2}
{|A\Omega_1|^2(1-b)\frac{\Gamma/2+i\delta_1}{\Gamma/2-i\delta_2}+
|\Omega_1|^2(1-bA^2)
+ i(\delta_1-\delta_2)(\Gamma/2+i\delta_1)}\right\} \, .
\end{eqnarray}

\subsection{Homogeneous broadening}
Consider first the case of ${\bf v}=0$.
The steady-state zero-order populations $\rho^{(0)}_{11}$ and
$\rho^{(0)}_{33}$ are governed by the equilibrium between the
excitation and relaxation processes in the absence of the probe field.
Obviously, these values can not contain structures with the width
less than $\Gamma$. Then, only the last two terms on the r.h.s. of
(\ref{final}) are responsible for the subnatural-width resonance on
the frequency offset $\delta_1-\delta_2$. If $\delta_1=0$, the sign
of the absorption resonance is determined by the sign of the
expression $(2b-1)\rho^{(0)}_{33}-\rho^{(0)}_{11}$ and, consequently,
depends on both the branching ratio constant and zero-order
populations. For example, in the absence of the spontaneous transfer
of the low-frequency coherence ($b=0$) the resonance is always
negative, that corresponds to EIT. In the opposite case of the closed
transition ($b=1$) the resonance is positive if
$\rho^{(0)}_{33} > \rho^{(0)}_{11}$, i.e. we have EIA (see in fig.2).
The position and the width of the EIA-resonance is determined by
the real and imagine parts of
the linear combination of the complex ground-level optical shifts:
$$
(1-b)\Delta\varepsilon_1-(1-bA^2)\Delta\varepsilon_3^{*} =
(1-b)\frac{|A\Omega_1|^2}{\delta_1-i\Gamma/2 }-
(1-bA^2)\frac{|\Omega_1|^2}{\delta_1+i\Gamma/2} \,.
$$
It is remarkable that the
coefficients of this combination depend on the branching ratio $b$.

\subsection{Doppler broadening}
In the case of atomic gas the EIA-resonance is a sum of structures
with different amlitudes, positions, and width. Here we consider the
result of such velocity averaging in one specific case. Let the
copropagating control and probe fields  have the approximately equal
Doppler shifts
$({\bf k}_1{\bf v})\approx({\bf k}_2{\bf v})\approx k v_z$.
Besides, we assume that the transition is closed $b=1$, and
the zero-order populations
$\rho^{(0)}_{11}=0$ and
$\rho^{(0)}_{33}=f_M({\bf v})$ is the Maxwell distribution. The
averaged optical coherence is expressed through the error function:
\begin{eqnarray} \label{averaging}
\langle \rho_{23} \rangle_v &=& -iB\Omega_2\left\{
V(\delta_2)+\frac{|A\Omega_1|^2}{|A\Omega_1|^2+i\Gamma
(\delta_1-\delta_2)-(\delta_1-\delta_2)^2}
\left[V(\delta_2)+V(-\delta_1+\frac{|B\Omega_1|^2}{\delta_1-\delta_2})
\right] \right\} \nonumber \\
V(x) &=& \frac{\sqrt{\pi}}{k \overline{v}}
\exp\left[\left(\frac{\Gamma/2-ix}{k \overline{v}}\right)^2\right]
\left(1+\mbox{erf}\left(\frac{ix-\Gamma/2}{k
\overline{v}}\right)\right) \,,
\end{eqnarray}
where $\delta_q =\omega_q-\omega_0$,
$\overline{v} = \sqrt{2 k_B T/M}$, and $V(x)$ is
the well-known Voight contour. In the case of large Doppler broadening
$k\overline{v} \gg \Gamma$  the probe field absorbtion
$\sim \mbox{Re}\{i\Omega^{*}_2\langle \rho_{23} \rangle_v\}$ as a
function of the frequency offset $\delta_1-\delta_2$ contains two
structures situated at $\delta_1-\delta_2=0$ with different width and
signs. One of them described by the Lorentzian
$|A\Omega_1|^2/(|A\Omega_1|^2+i\Gamma (\delta_1-\delta_2))$
has the width $|A\Omega_1|^2/\Gamma$ and gives the rising of
absorption. The last Voight function in
eq.(\ref{averaging}) $V(-\delta_1+|B\Omega_1|^2/(\delta_1-\delta_2))$
describes a very narrow dip in the absorption spectrum (see. in
fig.3.a). This structure with the width $|B\Omega_1|^2/(k
\overline{v})$ is the result of averaging. For an atom with given
velocity ${\bf v}$ the effective detuning is $\delta_1 - k v_z$ due
to the Doppler shift. As is seen from eq.(\ref{final}) and fig.2.b
the subnatural resonance is optically shifted with respect to the
point $\delta_1-\delta_2=0$. The sum of such shifted resonanes with
amplitudes and width depentent of $v_z$ gives a dip.  If
$\delta_1 \neq 0$, the absorption as a function of
$\delta_1-\delta_2$ becomes asymmetric (see in fig.3.b).

\section{Conclusion}
To conclude we note the four-state $N$-type interaction scheme can be
easily organized in real atomic systems. For example, let us consider
a closed $F_g=F \rightarrow F_e=F+1$ transition of the $D_2$ line of
an alkali atom interacting with the $\sigma_{+}$ polarized control
field. In the steady state all atoms are completely pumped into the
stretched states $|F_g,\,m_g=F\rangle$ and $|F_e,\,m_e=F+1\rangle$.
If the probe field has $\sigma_{-}$ polarization, then in the first
order on the probe field amplitude $\Omega_2$ we have $N$-atom with
the states $|1\rangle=|F_g,\,m_g=F-2\rangle$,
$|2\rangle=|F_e,\,m_e=F-1\rangle$, $|3\rangle=|F_g,m_g=F\rangle$,
and $|4\rangle=|F_e,\,m_e=F+1\rangle$.

\acknowledgments
We are grateful to authors of refs.\cite{EIA,EIA2} for 
stimulating discussions.

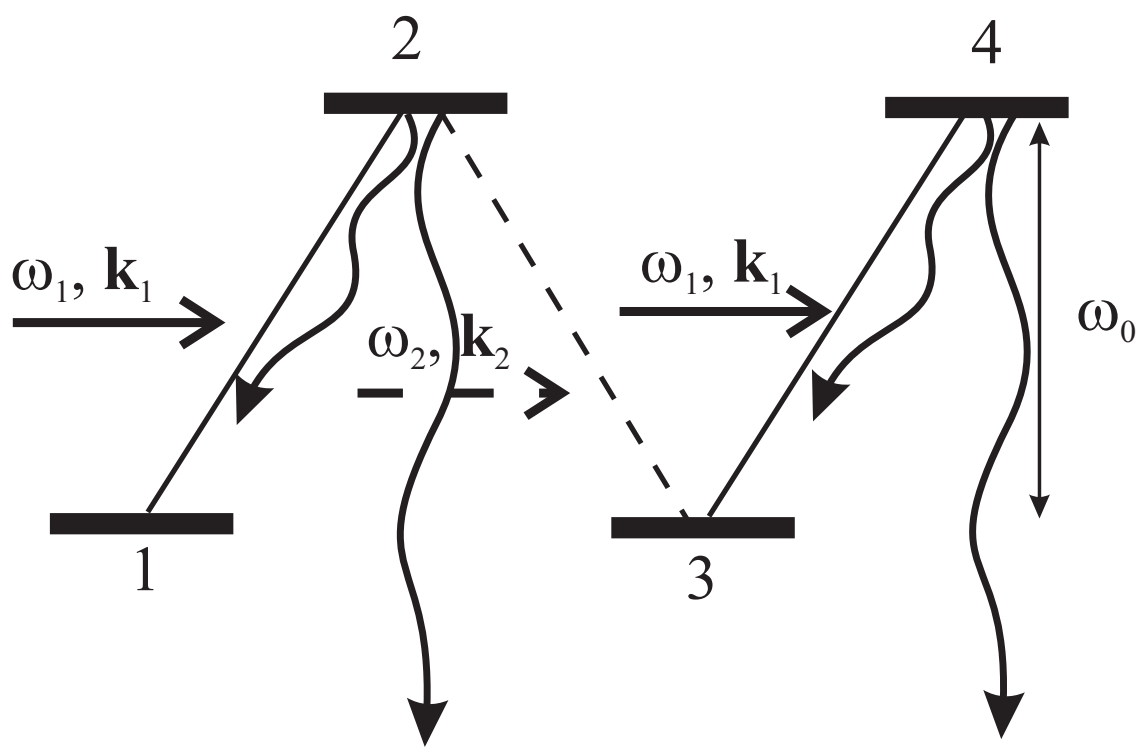
\begin{figure}
\caption{N-atom. The light-induced transitions are marked by solid
(control field) and dashed (probe field) lines. Wavy lines show two
possible channels of the spontaneous decay of the excited-level
coherence.}
\end{figure}

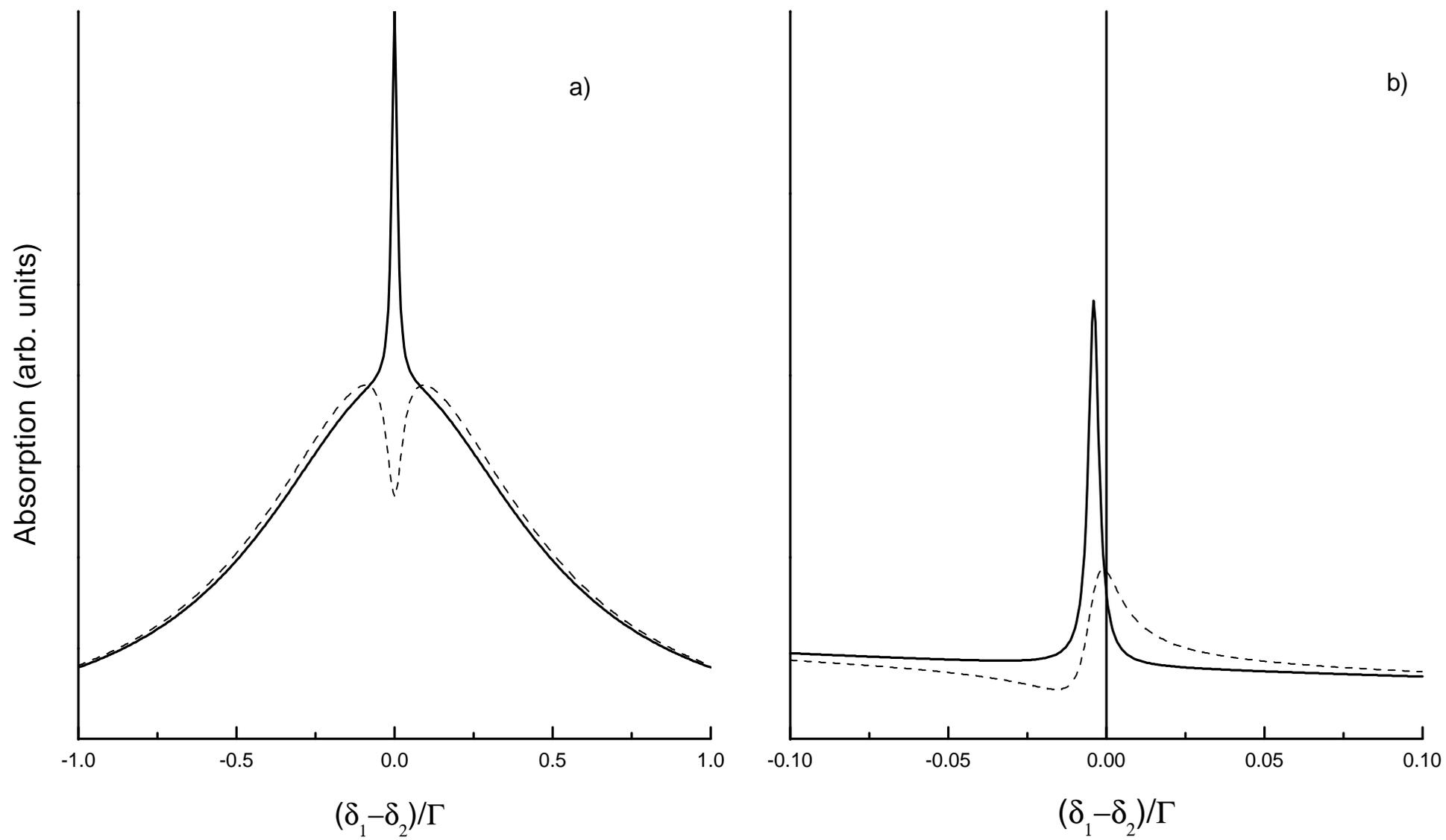
\begin{figure}
\caption{The probe field absorption versus the frequency offset
in the case of homogeneous broadening ${\bf v}=0$.
The control field detuning $\delta_1=0$ (a) and $\delta_1=\Gamma$
(b). Solid (dashed) curves correspond to the case of $b=1$ ($b=0$).
Other parameters are $A^2=B^2=1/2$,
$\Omega_1 = 0.1 \Gamma$, $\rho^{(0)}_{11}=0$
and $\rho^{(0)}_{33}=1$.}
\end{figure}

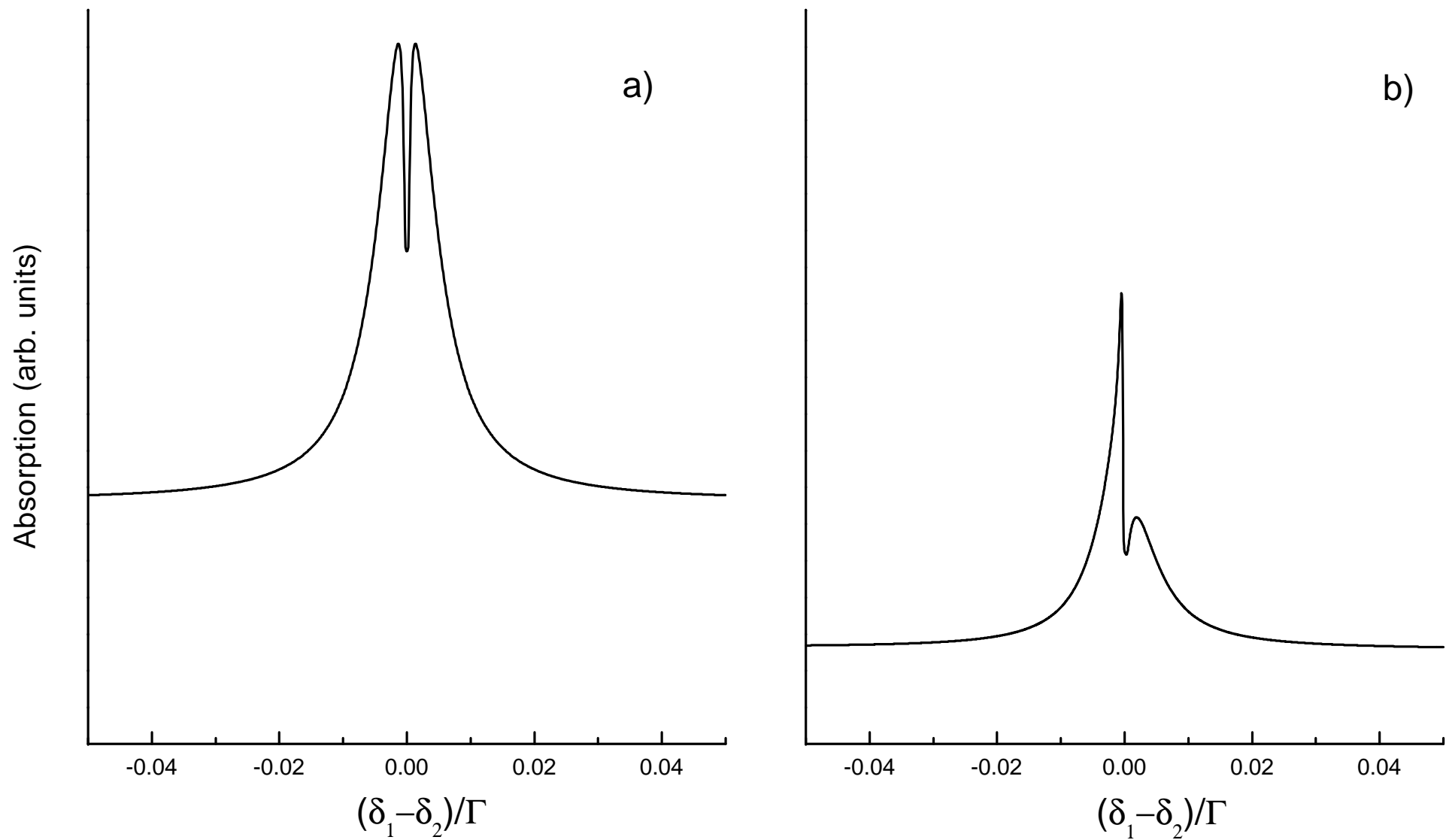
\begin{figure}
\caption{The probe field absorption versus the frequency offset
in the case of Doppler broadening $k\overline{v}= 10 \Gamma$. The
control field parameters: detuning $\delta_1=0$ (a) and
$\delta_1=10 \Gamma$ (b),
the Rabi frequency $\Omega_1 = 0.1 \Gamma$; and $A^2=B^2=1/2$.}
\end{figure}

\end{document}